\documentclass[fleqn,usenatbib]{mnras}

\usepackage{newtxtext,newtxmath}

\usepackage[T1]{fontenc}

\DeclareRobustCommand{\VAN}[3]{#2}
\let\VANthebibliography\thebibliography
\def\thebibliography{\DeclareRobustCommand{\VAN}[3]{##3}\VANthebibliography}

\usepackage{graphicx}	
\usepackage{amsmath}	

\newcommand{\rk}[1]{\textcolor{blue}{#1}}

\title[Reconstructing the Universe]{Reconstructing the Universe}

\author[Ryan E. Keeley et al.]{
Ryan E. Keeley,$^{1}$\thanks{E-mail: rkeeley@kasi.re.kr}
Arman Shafieloo,$^{1,2}$
Gong-Bo Zhao$,^{3,4}$
Jose Alberto Vazquez,$^{5}$
\newauthor
Hanwool Koo$^{1,2}$
\\
$^{1}$Korea Astronomy and Space Science Institute, Daejeon 34055, Korea\\ 
$^{2}$University of Science and Technology, Daejeon 34113, Korea\\
$^{3}$National Astronomy Observatories, Chinese Academy of Sciences, Beijing, 100012, P.R.China\\
$^{4}$University of Chinese Academy of Sciences, Beijing 100049, China\\
$^{5}$Instituto de Ciencias Fısicas, Universidad Nacional Autonoma de Mexico, Cuernavaca, Morelos, 62210, Mexico\\
}

\date{Accepted XXX. Received YYY; in original form ZZZ}

\pubyear{2020}

\begin{document}
\label{firstpage}
\pagerange{\pageref{firstpage}--\pageref{lastpage}}
\maketitle

\begin{abstract}
We test the mutual consistency between the baryon acoustic oscillation measurements from the eBOSS SDSS final release, as well as the Pantheon supernova compilation in a model independent fashion using Gaussian process regression.  We also test their joint consistency with the $\Lambda$CDM model, also in a model independent fashion.  We also use Gaussian process regression to reconstruct the expansion history that is preferred by these two datasets.  While this methodology finds no significant preference for model flexibility beyond $\Lambda$CDM, we are able to generate a number of reconstructed expansion histories that fit the data better than the best-fit $\Lambda$CDM model.  These example expansion histories may point the way towards modifications to $\Lambda$CDM.
We also constrain the parameters $\Omega_k$ and $H_0r_d$ both with $\Lambda$CDM and with Gaussian process regression.  We find that $H_0r_d =10030 \pm 130$ km/s and $\Omega_k = 0.05 \pm 0.10$ for $\Lambda$CDM and that $H_0r_d = 10040 \pm 140$ km/s and $\Omega_k =  0.02 \pm 0.20$ for the Gaussian process case.
\end{abstract}

\begin{keywords}
keyword1 -- keyword2 -- keyword3
\end{keywords}



\section{Introduction}
$\Lambda$CDM 
($\Lambda$ for a cosmological constant and CDM for cold dark matter) has emerged as the concordance model of cosmology.  This model explains a number of datasets well, at least individually.  In broad strokes, the $\Lambda$CDM model explains well the anisotropies in the cosmic microwave background, how those anisotropies cluster and grow into the observed large scale structure of the Universe, and how the expansion of the Universe accelerates at late times.

However, there have emerged a number of tensions in the $\Lambda$CDM parameters inferred by different datasets.  Most notably is the so-called ``$H_0$ tension'', which is a 4.4$\sigma$ discrepancy between the present-day expansion rate directly observed from low-redshift distances, including Cepheid anchoring of supernova distances (SN)~\citep{Riess2019}, strong lens time delay distances~\citep{H0LiCOW,keeleykai}, Tip of the Red Giant Branch~\citep{TRGB}, and that rate inferred from Planck measurements of the cosmic microwave background~\citep{Planck18Cosmo}.  There are a number of other tensions involving the growth of structure~\citep{KIDS450,KIDS1000}, and the inferred curvature~\citep{diValentino}.  Taken together, these may point towards a discrepancy between high and low-redshift physics~\citep{Keeley19GP2}.

The $H_0$ tension is primarily about the absolute scale of the lower end of the distance-redshift relation but the shape of this relation can point towards possible extension to $\Lambda$CDM that may explain the $H_0$ tension.  Two tracers of the shape of the distance-redshift relation are SN distances and the baryon acoustic oscillations (BAO) feature in the clustering of galaxies.
Jointly, the SN and BAO datasets are particularly relevant for this $H_0$ tension because they are anchored by the two datasets in question (Cepheids and CMB).  Indeed, if the $H_0$ from the Cepheids is taken to anchor the SN and $r_d$, the size of the Universe at the drag epoch, is taken to anchor the BAO, then the distances inferred from the two datasets are discrepant regardless of any cosmological interpretation of those distances~\citep{HubbleHunter}.  Thus the two datasets, even on their own and unanchored, can point the way towards what new physics might be needed to explain the $H_0$ tension.

While there have emerged a large number of explanations that have reconciled the Planck ``TT'' dataset and the Cepheid dataset, most have not been able to jointly explain every cosmological dataset.  Notably, explanations tend to fail explaining the polarization datasets~\citep{Neff,Rocknroll,EDE,EDE2}, or the large scale structure including the BAO~\citep{Keeley19GP2,GEDE,PEDE}, or the SN distances~\citep{GEDE,PEDE}. Rather than iterate through a possibly infinite number of nested extensions to $\Lambda$CDM or discrete alternative models, it can be more fruitful to use model independent methods.  That is, it is better to use data driven techniques to reconstruct the distance-redshift relation from the data directly. With the reconstructions, one can then build a model around what the data are trying to say.

In this paper, we first (Sec. 2) seek to use model independent methods to test that the SN and BAO distances are, in fact, consistent with each other and that they are jointly consistent with the $\Lambda$CDM model. In Sec. 3, we then reconstruct the expansion history of the Universe ($h(z)$) implied from these two datasets, as well as additional diagnostics that test the consistency with $\Lambda$CDM ($om(z)$ and $q(z)$).  We continue in Sec. 4 where we use our model independent methods to constrain the relative anchoring of the two SN and BAO distance datasets ($H_0 r_d$) as well as constrain the curvature ($\Omega_k$).  We compare the model independent results with those inferred by $\Lambda$CDM.  We discuss these results and conclude in Sec. 5.

\section{Consistency Tests}
In this section, we use Gaussian process (GP) regression, to perform model independent tests of the mutual consistency of the Pantheon~\citep{Pantheon} and SDSS~\citep{SDSSOverview} eBOSS~\citep{eBOSSOverview,eBOSSCosmo} BAO datasets and their joint consistency with the $\Lambda$CDM model.  It is important to perform consistency tests to answer whether there are any systematics which might hinder the ability to interpret these distances accurately.  Further, if the datasets have some certain systematic between them, any attempt to derive cosmological parameters from a joint inference of the two datasets will yield inaccurate and artificially precise results.  The inferred posteriors would be meaningless. Additionally, it is important to perform these kinds of tests in a model independent manner so that we can avoid making assumptions that we would want to test, thus making any outcome of the test more robust.

\subsection{Datasets}
Both the Pantheon~\citep{Pantheon} and SDSS BAO~\citep{eBOSSCosmo} datasets are unanchored so it is trivial to get the absolute scale of the datasets to agree. Thus we are effectively only testing if the shape of the inferred expansion histories are consistent.  In other words, these two datasets, on their own, will not be able to adjudicate which value of $H_0$ is correct, but it will be able to adjudicate the kinds of beyond-$\Lambda$CDM modifications can explain the tension.  If the two datasets are consistent with each other and with $\Lambda$CDM, then we can be more sure that the modification must occur outside the redshift range of the datasets.

The Pantheon SN dataset is composed of 1048 Type Ia SN between $z = 0.01$ and $z=2.3$.  SN are able to constrain cosmological distances because they are empirically \rk{assumed} to be standardizable candles.  That is, SN with the same light-curves, modulo the color and stretch of the SN and properties of the host galaxy, are thought to have the same intrinsic luminosity. This intrinsic luminosity, parametrized by $M_b$, is unknown, however, and degenerate with $H_0$.  Thus measuring the brightness of a SN can yield information about the relative distances of the SN but not their absolute distances.  SN constrain the shape of the expansion history but they are unanchored. 

The SDSS baryon acoustic oscillation (BAO) dataset measures the correlation function of galaxies. This correlation function contains a ``BAO feature'' which is an overdensity of power at the drag scale $r_d$, which is the size of the Universe at the drag epoch.  This feature arises from sound waves in the plasma of the early Universe.  This scale evolves along with the expansion of the Universe and so is encoded in the clustering of galaxies. Overdensities of galaxies are more likely to be located at a distance of $r_d$ apart.  

In a similar case to the SN, since $r_d$ is unknown the BAO datasets cannot be used alone to constrain the low redshift distances (or conversely, since the distances to these clustering galaxies is unknown, BAO datasets cannot be used alone to measure $r_d$), thus making the BAO datasets unanchored.  Like $M_b$, $r_d$ is another nuissance parameter to be marginalized over when comparing the SN and BAO distances

Because the distances to these clustered galaxies is unknown, the BAO datasets on their own to not constrain an absolute scale for the expansion history of the Universe.  As such, the BAO can constrain either the drag scale in dimensionless units of $h^{-1}$Mpc or the low-redshift distances relative to the drag scale $D_{V,M,H}/r_{\rm drag}$. Equivalently, the BAO datasets can be anchored either at high or low redshift with $r_{\rm drag}$ or $H_0$ respectively. 

The SDSS eBOSS final release measures the BAO feature in a variety of tracers over a variety of redshifts, with the Main Galaxy Sample at $z=0.15$~\citep{RossBAOMGS,HowlettBAOMGS} the BOSS Luminous Red Galaxy sample at redshifts $z = 0.38$ and $0.51$~\citep{SDSSLRGBeutler,SDSSLRGAlam} (the $z=0.61$ bin from this sample is merged into the eBOSS sample), the eBOSS Luminous Red Galaxy sample at redshift $z=0.70$~\citep{eBOSSCosmo},  the eBOSS Emission-line Galaxy sample at redshifts $z=0.85$~\citep{eBOSSCosmo}, the eBOSS quasar sample at redshift $z=1.48$~\citep{eBOSSCosmo}, and the BOSS/eBOSS Lyman$\alpha$ forest and quasar sample at $z=2.33$~\citep{SDSSLya,eBOSSLya}.

\subsection{Gaussian Process}

\begin{figure*}
    \centering
    \includegraphics[width=0.49\textwidth]{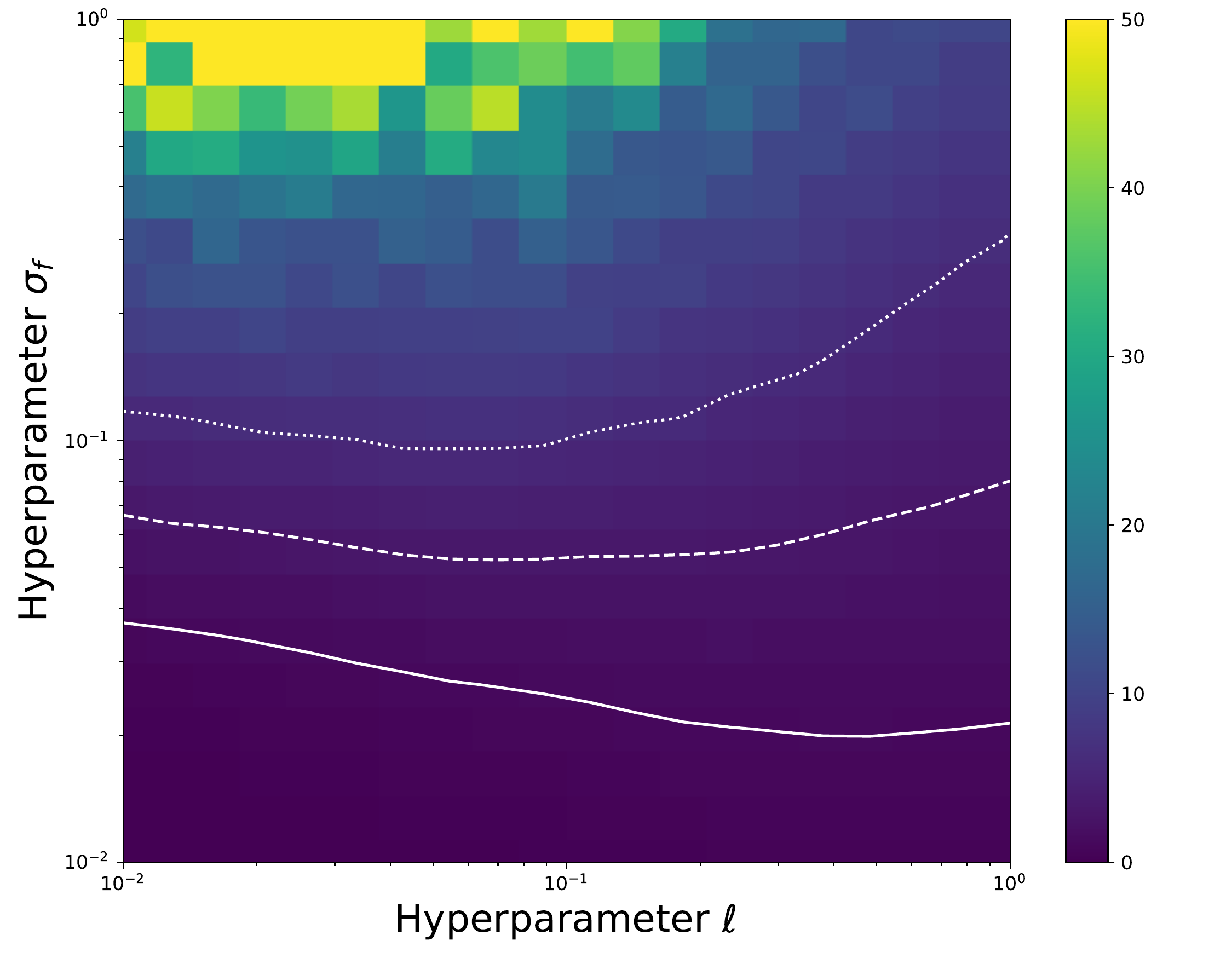}
        \includegraphics[width=0.49\textwidth]{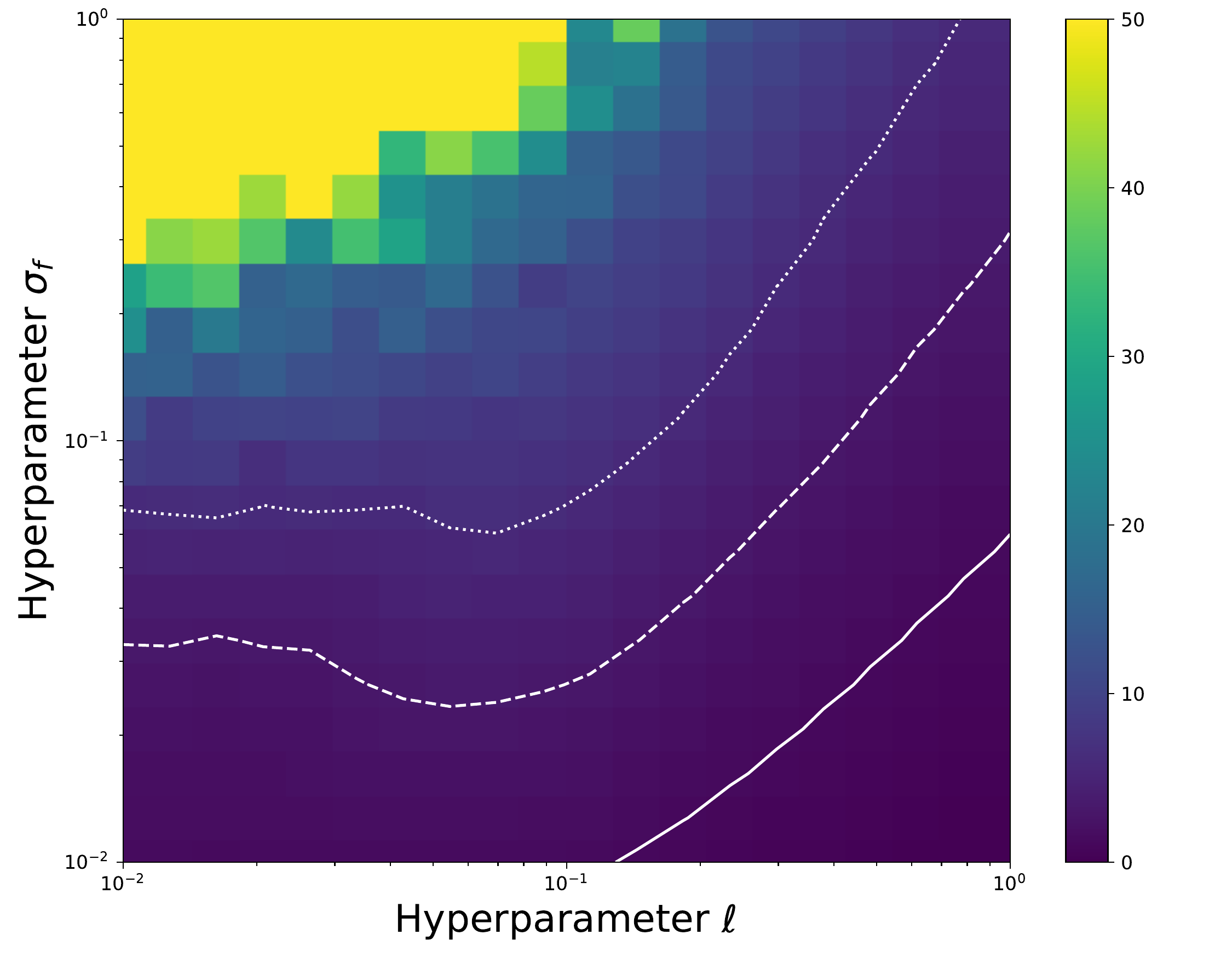}
            \includegraphics[width=0.5\textwidth]{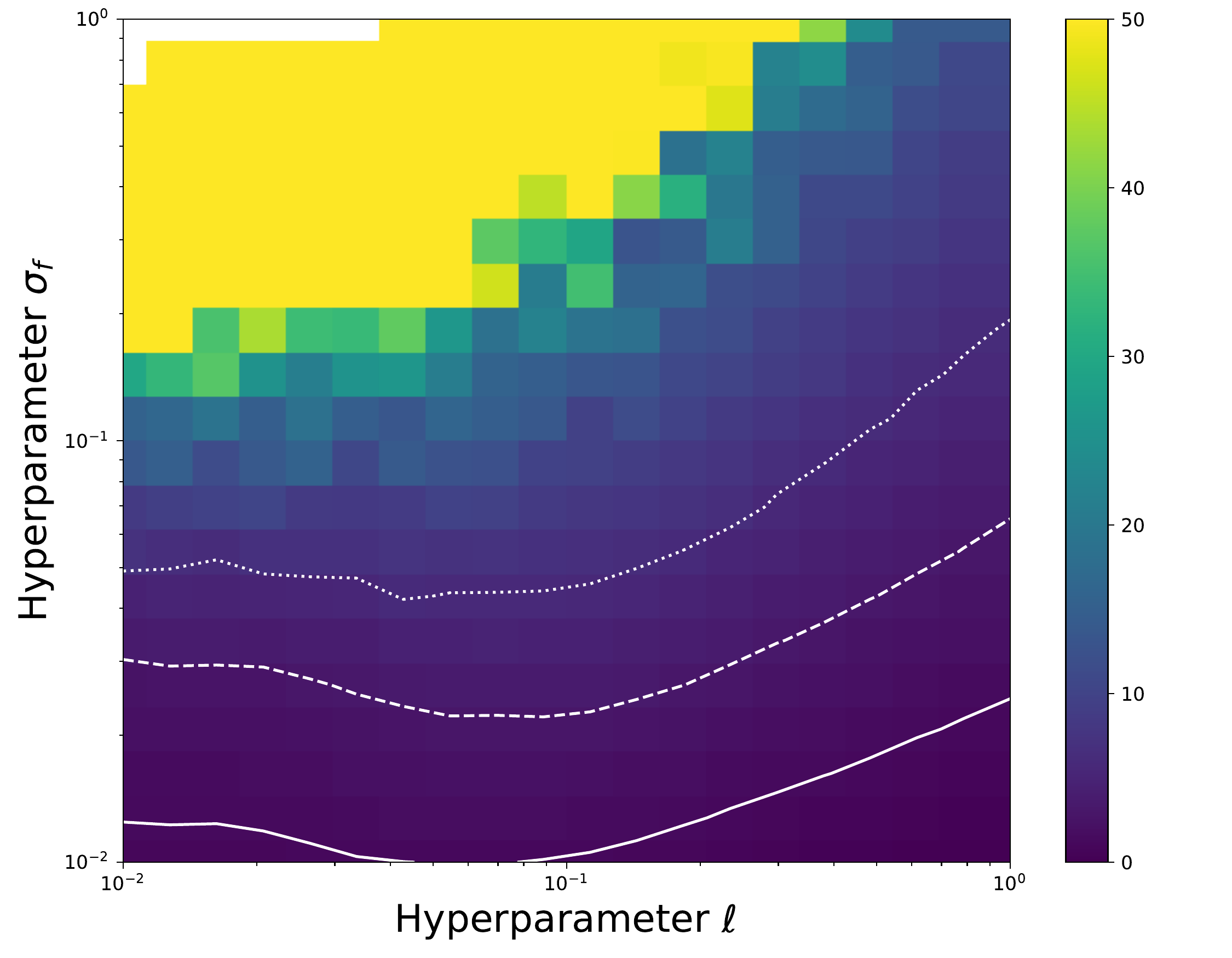}
    \caption{Hyperparameter posteriors for three different cases of consistency checks.  The solid, dashed, and dotted white lines correspond to the 1,2,3-$\sigma$ contours of the posterior, respectively. The top-left case corresponds to a GP inference of the SDSS data with a mean function taken from a best-fit GP sample to the SN data, the top-right is the same but switched: a GP inference of SN data with a mean function taken from a best-fit GP sample to the SDSS data.  The bottom case corresponds to a GP inference of both the SN and SDSS data with a mean function taken from the best-fit $\Lambda$CDM fit to the two datasets.}
    \label{fig:kernel}
\end{figure*}

A GP is an infinite collection of correlated random variables characterized by a covariance function~\citep{Rasmussen}.  Where a Gaussian distribution draws a single number, a GP generalizes this concept and draws a function.  So, in a sense, GP can be thought of as a sampling method but instead of sampling over a finite dimensional parameter space as in Markov chain Monte Carlo (MCMC), GP samples an infinite dimensional function space.  

GPs generally can take an input called a mean function, a function about which the random fluctuations of the GP varies. Therefore, the draws of a GP can be thought of as hyperfunctions (not the holomorphic variety, just a generalization of a hyperparameter, i.e. a function to be marginalized over) and thus GP can be used as a method to look for deviations away from this mean function.  To test $\Lambda$CDM, then we can choose it as the mean function for our GP inferences.

Using GPs to perform a regression utilizes both these understandings of GP.  We essentially take a GP as the prior in a Bayesian analysis.
We essentially sample the prior for a Bayesian analysis using a GP.
When we want to sample the prior for a Bayesian analysis, we use a GP as that prior.  In proper Bayesian fashion, we marginalize over this family of hyperfuntions drawn from the GP prior.  Each of these hyperfunction samples is weighted by their likelihood, by how well they fit the data.  Histogramming each of these weighted hyperfunctions then allows us to calculate the posterior, not in terms of a set of parameters, but directly in terms of the reconstructed cosmological functions, for instance, the expansion history $H(z)$.  We can summarize the methodology up to this point with the equation via Bayes' theorem,
\begin{equation}
    P(H(z)|D) = \int d \phi_{GP} \mathcal{L}(D|H(z,\phi_{GP})) P(\phi_{GP}) / P(D),
\end{equation}
where $\phi_{GP}$ is the family of hyperfunctions from the GP, $D$ is the data, $P(H(z)|D)$ is the posterior, $\mathcal{L}(D|H(z,\phi_{GP}))$ is the likelihood, $P(\phi_{GP})$ is the prior and $P(D)$ is the evidence.  The hyperfunctions are related to the expansion history by the following formula,
\begin{equation}
    H(z) = H_{mf}(z) \exp(\phi(z)) , 
\end{equation}
where $H_{mf}(z)$ is the expansion history of the mean function.
From the posterior, we can calculate quantities like the 68\% and 95\% confidence levels (CL) for the value of $H(z)$ at any particular redshift.  Joining the CLs for various redshifts allows us to generate the ``band'' plots, which if the data significantly prefer some amount of evolution in the expansion history, relative to the mean function, then it will show up in these plots.  

To be concrete about some of the details of GP, as mentioned previously, a GP is characterized by a covariance function.  This covariance function can be quite general so long as it satisfies some general properties like being symmetric and positive semidefinite. We specifically use a squared-exponential covariance matrix with the following form,
\begin{equation}
\langle \phi(s_1) \phi(s_2) \rangle = \sigma_f^2 \, e^{ -(s_1 - s_2)^2/(2\ell^2)}  \ ,
\end{equation}
where our evolution variable is $s(z) = \log(1+z)/\log(1+z_{\rm max})$.  We take $z_{\rm max}=3$.  Importantly, the covariance function is characterized by two hyperparameters. $\sigma_f$  determines the heights of the random fluctuations of the GP, i.e. the scale of the deviations away from the mean function.  If the data prefer additional information or flexibility beyond the input mean function, they they will pick out a value for $\sigma_f$ above zero.
$\ell$ which determines the length of the random fluctuations.  In simple terms, $1/\ell$ is roughly the number of independent random fluctuations in the range.  So GP samples with large $\ell$ and large $\sigma_f$ would have a few large deviations while small $\ell$ and small $\sigma_f$ would have many smaller deviations.  Because these hyperparameters encode information about the inferred expansion histories, the hyperparameters must be fit for and cannot be assumed.

One might be concerned that this sort of analysis is prone to overfitting.  If $\ell$ is small, say around $\ell \sim 0.001$, that is, in effect, $\sim 1000$ degrees of freedom and the worry is that this analysis could easily achieve an arbitrarily good $\chi^2$ value. It is possible for GP to generate a sample that has this feature, however, for a GP with $\ell \sim 0.001$, the variety of functions that it can produce is vast and so it would be rare for the GP to actually generate this hypothetical example.  In other words, because we marginalize over the space of possible functions, we avoid overfitting.  Bayesian analyses are typically safe from the overfitting problem.

It is these hyperparameters that can be used for testing whether the data is consistent with the mean function of the GP regression~\citep{ShafKimLind2012,ShafKimLind2013,AghHamShaf2017,BenShafKimLind2019,KeeleyGP3}.  Since $\sigma_f$ determines the size of the deviations away from the mean function, the test of the consistency of the mean function with the data amounts to testing if $\sigma_f$ is consistent with 0.  So in effect, we calculate the posterior for the hyperparameters $\sigma_f$ and $\ell$ and see where CLs end up.  This GP code is based in \texttt{gphist}~\citep{gphistdoi}, which is first introduced in \cite{KeeleyGP1} and later refined in \cite{Keeley19GP2} and \cite{KeeleyGP3}.
 
\subsection{Testing $\Lambda$CDM}
Now that we have a general methodology to test whether the mean function of a GP is consistent with the data, we can test if the two datasets are jointly consistent with the $\Lambda$CDM model.  This is achieved simply by fitting $\Lambda$CDM to the two datasets, finding which parameters fit them best, and then using the expansion history of this best-fit $\Lambda$CDM model as the mean function.  Thus if posterior of $\sigma_f$ is consistent with 0, then $\Lambda$CDM is consistent with the two datasets jointly. We see the results of this test in the bottom panel of Fig.~\ref{fig:kernel}.  There the posterior is shown with the color corresponding to $-\log (P(\sigma_f, \ell | D))$ and the 68\%, 95\% 99.7\% CLs.  We see that $\sigma_f = 0$ and thus $\Lambda$CDM is consistent with the two datasets.

\subsection{Testing mutual consistency}
In order to test the mutual consistency of the two datasets we can choose the mean function to be a GP reconstruction of one of the datasets to serve as the mean function for a GP reconstruction of the other.  Schematically, we start with the best-fit $\Lambda$CDM expansion history to the SDSS BAO dataset, for example, which we use as a mean function for a GP regression of that same dataset.  We then use the median GP reconstruction of the SDSS BAO dataset as a mean function for a GP regression of the Pantheon dataset.  The posterior of the hyperparameters for this second GP regression is what is shown in the upper panels of Fig.~\ref{fig:kernel}.  The top-left panel of that figure corresponds the GP regression of the SDSS BAO dataset with a GP reconstruction from the Pantheon dataset as a mean function.  The top-right panel is the reverse, a GP regression of the Pantheon dataset with a mean function taken from the GP reconstruction from the SDSS BAO dataset.  In both cases, we see that the posteriors of the hyperparameters are consistent with $\sigma_f = 0$ and thus the data prefer no additional information beyond the mean function.  The distances inferred from the SDSS BOSS and Pantheon datasets are consistent.

\section{Reconstructions}
\begin{figure*}
\centering
\includegraphics[width=0.49\textwidth]{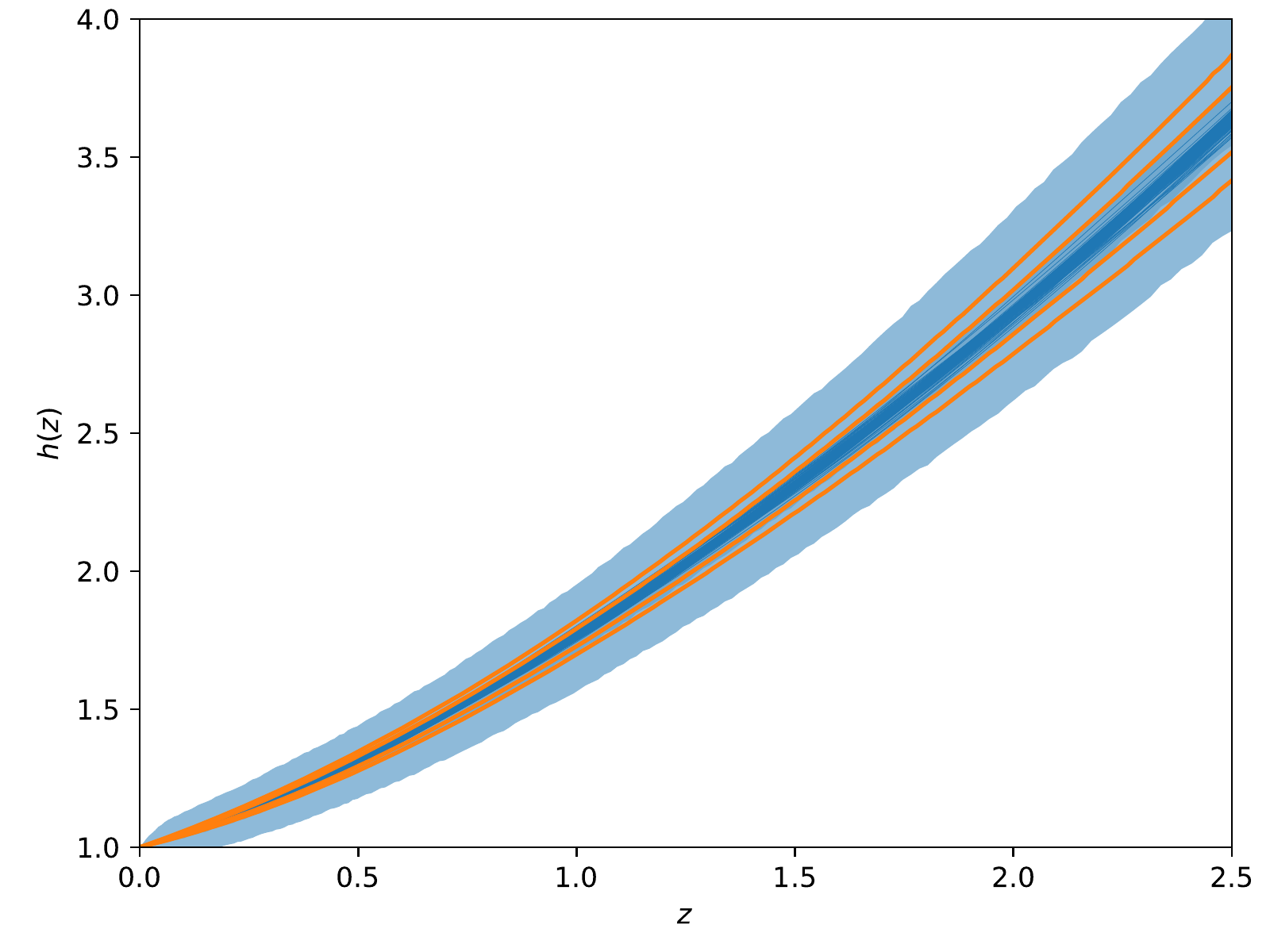}
\includegraphics[width=0.49\textwidth]{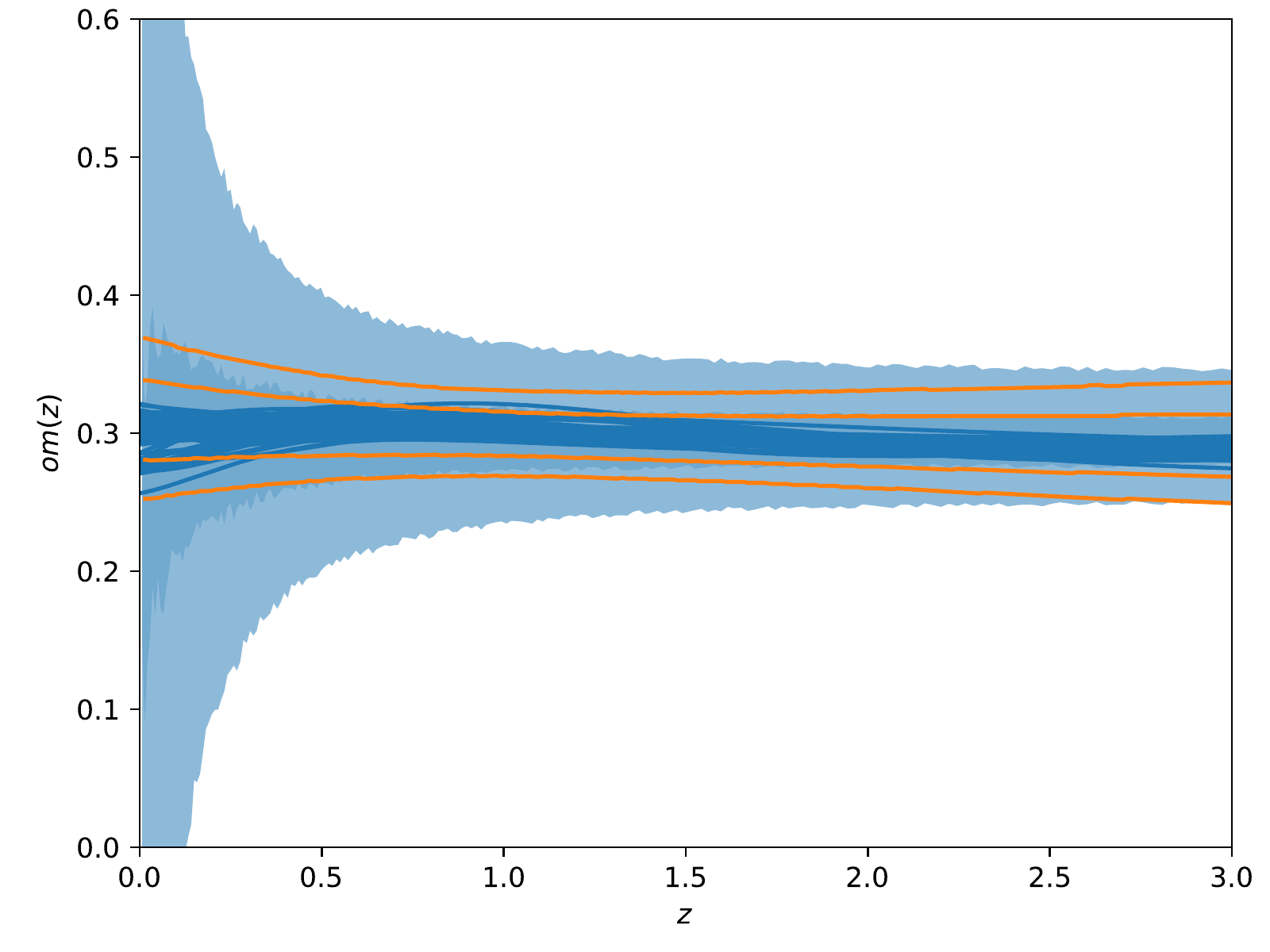}
\includegraphics[width=0.49\textwidth]{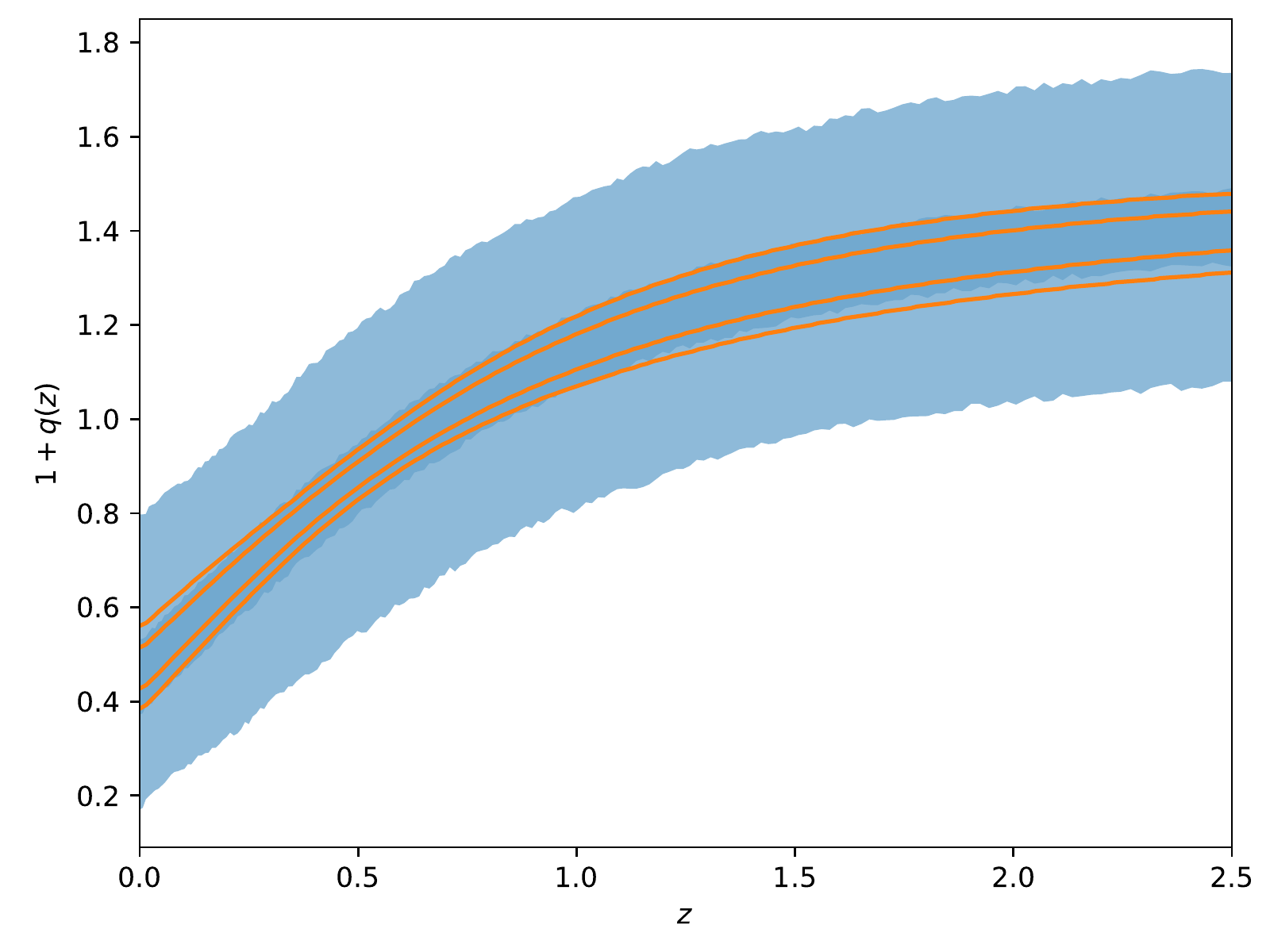}
\caption{GP reconstructions of $h(Z)$, $q(z)$, and $om(z)$.  The blue bands represent the 68\% and 95\% confidence levels for the GP reconstructions and the darkest blue lines represent example GP samples that fit the data better than the best-fit $\Lambda$CDM model.  The orange lines enclose the 68\% and 95\% confidence levels for the $\Lambda$CDM case.}
\label{fig:recon}
\end{figure*}

In this section, we present reconstructions of various parameters of the Universe's expansion history that are independent of the absolute scale of the expansion history, including $h(z) = H(z)/H_0$, the deceleration parameter $q(z)$, and the om diagnostic $om(z)$.
The deceleration parameter is given by the following formula,
\begin{equation}
  q(z) = -\frac{\dddot{a}a}{\ddot{a}^2} = -1 + \frac{d\log H(z) }{ d\log(1+z)  } ,
\end{equation}
and the om diagnostic is given by
\begin{equation}
    om(z) = \frac{h(z)^2 - 1}{(1+z)^3 -1} .
\end{equation}

Each of these parameters are potentially useful for testing whether the GP reconstructions are consistent with $\Lambda$CDM or if some evolution is preferred.
The dimensionless expansion histpory, $h(z)$, is the most obvious and direct quantity to reconstruct from the data, however, its interpretation is more uncertain because the variation within $\Lambda$CDM can look categorically similar to the variation within models of evolving dark energy.  Specific functions of $h(z)$, however, can make more robust tests of $\Lambda$CDM or not $\Lambda$CDM.

For instance, $om(z)$ should be constant and equal to the matter density $\Omega_{\rm m}$ if $\Lambda$CDM correctly described the low-redshift distances.  Any evolution in the expansion history, for example from an evolving dark energy, would show up as an evolution in this parameter.

Similarly, for $\Lambda$CDM $1+q(z) = \frac{3}{2} \frac{\Omega_m(1+z)^3 }{ \Omega_{\rm crit}(z)}$, so at large redshifts it is equal to $3/2$ and then transitions to $3/2 \ \Omega_m$ by $z=0$.  Since this parameter is, effectively, the second derivative of the data, using a model independent method to constrain this quantity will necessarily give uncertain and noisy results.

The results of these reconstructions are shown in Fig.~\ref{fig:recon}. The lighter blue shaded regions correspond to the 68\% and 95\% CLs of the GP regression.  These quantities are calculated by taking each GP reconstruction, calculating $h(z)$, $om(z)$, and $q(z)$ for each, then, for each redshift, grab the quantity at that redshift, and then make a histogram over all the different reconstructions, weighted by their likelihoods.  We use these histograms to calculate the 68\% and 95\% CLs.

In orange, we show the corresponding CLs but for the case where $\Lambda$CDM is assumed.  The notable feature here is that the $\Lambda$CDM case yields tighter constraints on these functions of the expansion history.  This feature is expected, since , by construction, the GP case is more agnostic about the expansion history.

On top of these bands, we plot specific GP reconstructions that have a better $\chi^2$ than the best-fit $\Lambda$CDM.  The reconstructions of $om(z)$ have some noticeable, though rare, evolution towards low redshift, thought this is the region where the data is least constraining on this quantity.

From these reconstructions, we can see that the Pantheon and SDSS BAO datasets prefer no significant evolution with respect to $\Lambda$CDM model, and show a non-exhaustive set of example expansion histories that happen to fit these datasets better then the best-fit $\Lambda$CDM.

However, there do exist a number of GP reconstructions that fit the data better than the best-fit $\Lambda$CDM model.  An examination of the GP hyperparameters that generated these reconstructions can give insight into what features of the data the reconstructions are fitting better than $\Lambda$CDM.  Most of these better than $\Lambda$CDM reconstructions have large $\ell$ values while only having a better fit by $\Delta \chi ^2 \sim 0.1 - 1.0$.  The GP reconstructions that have the best fit to the data ($\Delta \chi^2 = 6$) also have the smallest values of $\ell$, which, in some sense, translates to a large number of degrees of freedom, so these reconstructions have rapidly varying $h(z)$.  The vary in such a way that $h(z)$ is not monotonically increasing with redshift, indicating for that reconstruction, the inferred dark energy density would have to be negative at some point.  It might be reasonable to reject any reconstruction that would have a negative dark energy density at any point on purely a priori grounds, but it is still interesting to see what sort of expansion history is needed to fit the data better than $\Lambda$CDM.  Taken together, each of these discussed features point towards the conclusion that these reconstructions that fit the data better than $\Lambda$CDM are merely over-fitting the noise in the data.

You can, of course, hack your way to a significant model by taking these reconstructions and contriving a model that generically predicts the reconstructions over the entirety of its parameter space.  This feature of model-fitting methods demonstrates another aspect of why model-independent methods are useful; you cannot hack your way to a significant result.

\section{Anchors and Curvature}
In this section, we use the SDSS BAO measurements alongside the Pantheon SN distances to constrain $H_0r_d$.  This parameter is the relative anchor of these two unanchored datasets.  The SN constrain unanchored distances, or similarly, they constrain $E(z) = H(z)/H_0$.  The individual BAO constraints measure $H(z)r_d$ or $D_M/r_d$ at various redshifts.  However, to get $H_0r_d$, one still needs to make assumptions to project down to $z=0$.  This is where the SN enter, since it is in this region where they have the greatest constraining power. Thus combining the two dataset can yield robust and tight constraints on $H_0r_d$, even with model-independent methods.

It is trivial to calculate this in the $\Lambda$CDM case. The low redshift expansion history and distances are simply given by the typical parameters $H_0$, $\Omega_m$ and $\Omega_k$.  We also fit for $M_b$ and $r_d$ to calibrate the distances from the Pantheon and SDSS BAO datasets.  Where normally, within $\Lambda$CDM, $r_d$ is a derived parameter which depends on the other parameters of the background model ($H_0$, $\Omega_b$, $\Omega_{cdm}$, $\Omega_k$), in this analysis, because we are only fitting low-redshift distances, we seek to remain agnostic about any potential beyond-$\Lambda$CDM modifications which might affect $r_d$.  Therefore, we treat $r_d$ as an independent parameter and fit for it independently of the other parameters.  In summary, we vary the five mentioned parameters ($H_0$, $\Omega_m$, $\Omega_k$, $M_b$, and $r_d$) and use MCMC to calculate the posterior of these parameters.  Thus it is relatively trivial to express the constraint in terms of $H_0r_d$, the samples of $H_0r_d$ are simply the multiplication of the samples of $H_0$ and $r_d$.

Each sample from the GP is a randomly generated $H(z)$, so $H_0$ in this case is simply $H(z=0)$.  For the GP case, we also treat $M_b$ and $r_d$ as nuisance parameters, so we also fit for these parameters alongside the expansion histories generated from the GP.  Thus, again, the samples of $H_0r_d$ are simply the multiplication of the samples of $H(z=0)$ and $r_d$.  Since the BAO dataset constrains both the distances $D_M(z)$ and the expansion rate $H(z)$, the GP can constrain the curvature, since 
\begin{equation}
    D_M(z) =  \Omega_k^{-1/2} \sinh\left[ \Omega_k^{1/2} c \int^z_0 dz'/H(z') \right].
\end{equation}
The GP case cannot constrain the matter density in a similar way since for any $\Omega_m$, I can then choose $w(z)$ to get the $H(z)$ needed for the GP reconstruction.

For the $\Lambda$CDM case, we find that $H_0r_d = 10030 \pm 130$ km/s, $\Omega_m = 0.28 \pm 0.04$, and $\Omega_k = 0.05 \pm 0.10$ and for the GP case, we find that that $H_0r_d = 10040 \pm 140$ km/s and $\Omega_k = 0.02 \pm 0.20$. 

\begin{figure}
    \centering
    \includegraphics[width=\columnwidth]{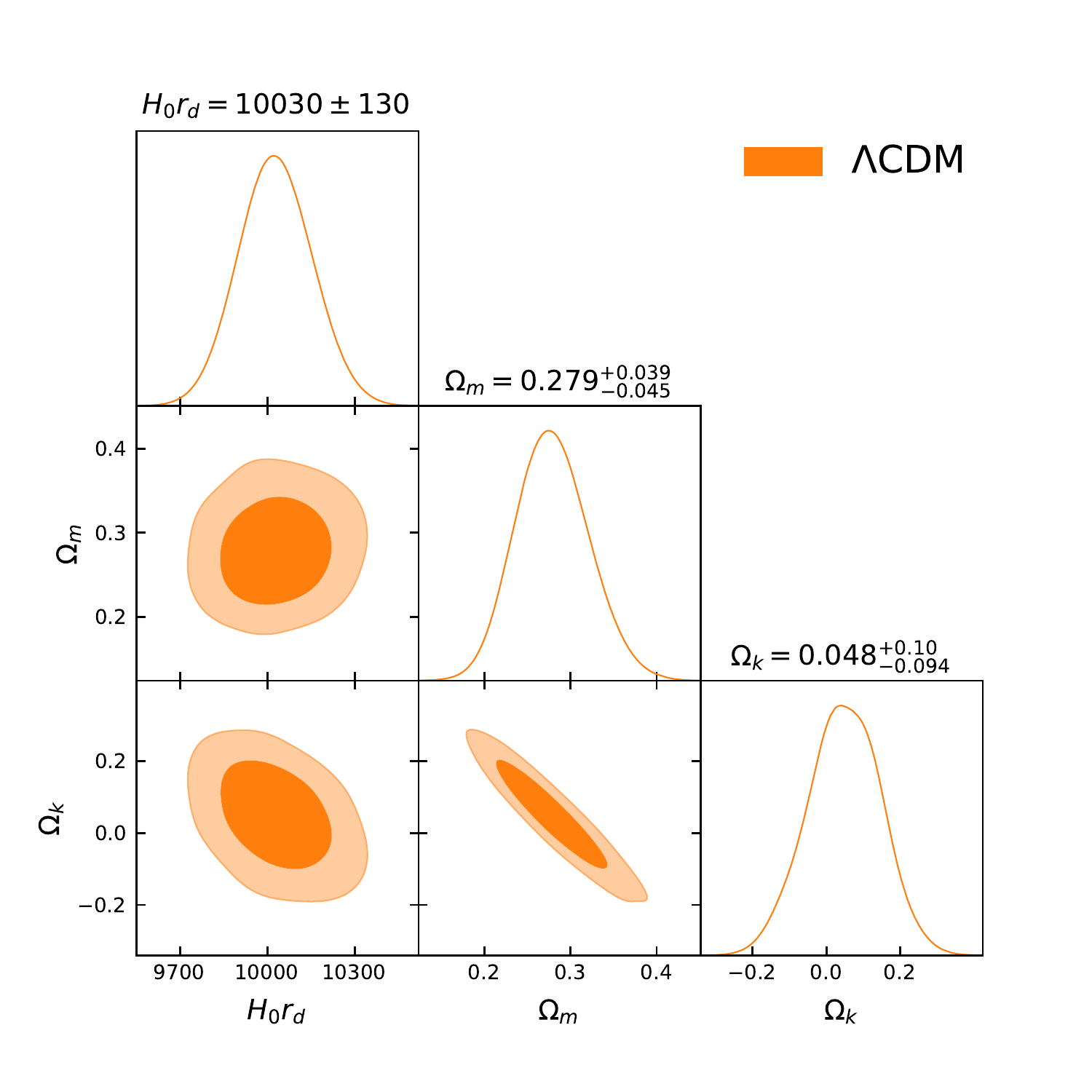}
    \caption{Posteriors of the $\Lambda$CDM parameters.}
    \label{fig:lcdm}
\end{figure}

\begin{table}
	\centering
	\caption{Best-fit $\chi^2$ values for the $\Lambda$CDM and GP cases.}
	\label{tab:example_table}
	\begin{tabular}{lccr} 
		\hline
		 & SN & SDSS & SN+SDSS \\
		\hline
		$\Lambda$CDM & 1027 & 7 & 1034\\
		GP & 1023 & 5 & 1028\\
		
		\hline
	\end{tabular}
\end{table}

\section{Discussion and Conclusions}
These results can be used to constrain the low redshift expansion history and so SN and BAO can adjudicate low-redshift explanations for the H0 tension, e.g. curvature or evolving dark energy. 

These results can also be used to constrain any potential evolution of the SN.  It has been claimed~\citep{kim-etal19} that SN might not show any evidence for an accelerating Universe or for dark energy.  The claim instead is that SN evolve with redshift, that either the light-curve calibration with SALT-2 or the Tripp formula would need extra freedom to account for properties of the host galaxy.  Our methodology implicitly constrains these ideas.  Since we found that both the reconstructed expansion histories from the SN and from the BAO are consistent with each other and with $\Lambda$CDM, any astrophysical evolution would break this consistency and thus be disfavored.

This shows that the inference of dark energy is dependent on not just the SN dataset; the BAO dataset confirms this as well.

In this paper, we use GP regression to show that the SN and BAO datasets are consistent with each other and with $\Lambda$CDM.  Further, we reconstruct dimensionless functions of the expansion history of the Universe, $h(z)$, $q(z)$, $om(z)$.  This allows us to visually inspect the consistency between these datasets and the $\Lambda$CDM model and to demonstrate that there is still some flexibility allowed at low redshifts. Finally, we calculate $\Lambda$CDM posteriors from the two datasets finding that $H_0r_d = 10030 \pm 130$ km/s, $\Omega_m = 0.28 \pm 0.04$, and $\Omega_k = 0.05 \pm 0.10$. We also constrain $H_0 r_d$ and $\Omega_k$ using our non-parametric GP method finding that $H_0r_d = 10040 \pm 140$ km/s and $\Omega_k = 0.02 \pm 0.20$.

\section*{Acknowledgements}
This work was supported by the high performance computing clusters Seondeok at the Korea Astronomy and Space Science Institute. A.~S. would like to acknowledge the support of the Korea Institute for Advanced Study (KIAS) grant funded by the government of Korea. B.~L. would like to acknowledge the support of the National Research Foundation of Korea (NRF-2019R1I1A1A01063740). G.B.Z. is supported by the National Key Basic Research and De- velopment Program of China (No. 2018YFA0404503), a grant of CAS Interdisciplinary Innovation Team, and NSFC Grants 11925303, 11720101004, 11673025 and 11890691. 

Funding for the Sloan Digital Sky Survey IV has been provided by the Alfred P. Sloan Foundation, the U.S. Department of Energy Office of Science, and the Participating Institutions. SDSS acknowledges support and resources from the Center for High-Performance Computing at the University of Utah. The SDSS web site is www.sdss.org.

SDSS is managed by the Astrophysical Research Consortium for the 
Participating Institutions of the SDSS Collaboration including the 
Brazilian Participation Group, the Carnegie Institution for Science, 
Carnegie Mellon University, 
Center for Astrophysics Harvard \& Smithsonian (CfA), 
the Chilean Participation Group, 
the French Participation Group, 
Instituto de Astrofísica de Canarias, 
The Johns Hopkins University, 
Kavli Institute for the Physics and Mathematics of the Universe (IPMU) 
/ University of Tokyo, 
the Korean Participation Group, 
Lawrence Berkeley National Laboratory, 
Leibniz Institut für Astrophysik Potsdam (AIP), 
Max-Planck-Institut für Astronomie (MPIA Heidelberg), 
Max-Planck-Institut für Astrophysik (MPA Garching), Max-Planck-Institut für Extraterrestrische Physik (MPE), National Astronomical Observatories of China, New Mexico State University, New York University, University of Notre Dame, Observatório Nacional / MCTI, The Ohio State University, Pennsylvania State University, Shanghai Astronomical Observatory, United Kingdom Participation Group, Universidad Nacional Autónoma de México, University of Arizona, University of Colorado Boulder, University of Oxford, University of Portsmouth, University of Utah, University of Virginia, University of Washington, University of Wisconsin, Vanderbilt University, and Yale University.



\bibliographystyle{mnras}
\bibliography{example}

\begin{thebibliography}{}
\makeatletter
\relax
\def\mn@urlcharsother{\let\do\@makeother \do\$\do\&\do\#\do\^\do\_\do\%\do\~}
\def\mn@doi{\begingroup\mn@urlcharsother \@ifnextchar [ {\mn@doi@}
  {\mn@doi@[]}}
\def\mn@doi@[#1]#2{\def\@tempa{#1}\ifx\@tempa\@empty \href
  {http://dx.doi.org/#2} {doi:#2}\else \href {http://dx.doi.org/#2} {#1}\fi
  \endgroup}
\def\mn@eprint#1#2{\mn@eprint@#1:#2::\@nil}
\def\mn@eprint@arXiv#1{\href {http://arxiv.org/abs/#1} {{\tt arXiv:#1}}}
\def\mn@eprint@dblp#1{\href {http://dblp.uni-trier.de/rec/bibtex/#1.xml}
  {dblp:#1}}
\def\mn@eprint@#1:#2:#3:#4\@nil{\def\@tempa {#1}\def\@tempb {#2}\def\@tempc
  {#3}\ifx \@tempc \@empty \let \@tempc \@tempb \let \@tempb \@tempa \fi \ifx
  \@tempb \@empty \def\@tempb {arXiv}\fi \@ifundefined
  {mn@eprint@\@tempb}{\@tempb:\@tempc}{\expandafter \expandafter \csname
  mn@eprint@\@tempb\endcsname \expandafter{\@tempc}}}

\bibitem[\protect\citeauthoryear{{Aghamousa}, {Hamann}  \&
  {Shafieloo}}{{Aghamousa} et~al.}{2017}]{AghHamShaf2017}
{Aghamousa} A.,  {Hamann} J.,   {Shafieloo} A.,  2017, \mn@doi [Journal of
  Cosmology and Astro-Particle Physics] {10.1088/1475-7516/2017/09/031}, \href
  {https://ui.adsabs.harvard.edu/abs/2017JCAP...09..031A} {2017, 031}

\bibitem[\protect\citeauthoryear{{Agrawal}, {Cyr-Racine}, {Pinner}  \&
  {Randall}}{{Agrawal} et~al.}{2019}]{Rocknroll}
{Agrawal} P.,  {Cyr-Racine} F.-Y.,  {Pinner} D.,   {Randall} L.,  2019, arXiv
  e-prints, \href {https://ui.adsabs.harvard.edu/abs/2019arXiv190401016A} {p.
  arXiv:1904.01016}

\bibitem[\protect\citeauthoryear{{Alam} et~al.,}{{Alam}
  et~al.}{2017}]{SDSSLRGAlam}
{Alam} S.,  et~al., 2017, \mn@doi [\mnras] {10.1093/mnras/stx721}, \href
  {https://ui.adsabs.harvard.edu/abs/2017MNRAS.470.2617A} {470, 2617}

\bibitem[\protect\citeauthoryear{{Bernal}, {Verde}  \& {Riess}}{{Bernal}
  et~al.}{2016}]{Neff}
{Bernal} J.~L.,  {Verde} L.,   {Riess} A.~G.,  2016, \mn@doi [\jcap]
  {10.1088/1475-7516/2016/10/019}, \href
  {https://ui.adsabs.harvard.edu/abs/2016JCAP...10..019B} {2016, 019}

\bibitem[\protect\citeauthoryear{{Beutler} et~al.,}{{Beutler}
  et~al.}{2017}]{SDSSLRGBeutler}
{Beutler} F.,  et~al., 2017, \mn@doi [\mnras] {10.1093/mnras/stw2373}, \href
  {https://ui.adsabs.harvard.edu/abs/2017MNRAS.464.3409B} {464, 3409}

\bibitem[\protect\citeauthoryear{{Blanton} et~al.,}{{Blanton}
  et~al.}{2017}]{SDSSOverview}
{Blanton} M.~R.,  et~al., 2017, \mn@doi [\aj] {10.3847/1538-3881/aa7567}, \href
  {https://ui.adsabs.harvard.edu/abs/2017AJ....154...28B} {154, 28}

\bibitem[\protect\citeauthoryear{{Dawson} et~al.,}{{Dawson}
  et~al.}{2016}]{eBOSSOverview}
{Dawson} K.~S.,  et~al., 2016, \mn@doi [\aj] {10.3847/0004-6256/151/2/44},
  \href {https://ui.adsabs.harvard.edu/abs/2016AJ....151...44D} {151, 44}

\bibitem[\protect\citeauthoryear{{Di Valentino}, {Melchiorri}  \& {Silk}}{{Di
  Valentino} et~al.}{2020}]{diValentino}
{Di Valentino} E.,  {Melchiorri} A.,   {Silk} J.,  2020, \mn@doi [Nature
  Astronomy] {10.1038/s41550-019-0906-9}, \href
  {https://ui.adsabs.harvard.edu/abs/2020NatAs...4..196D} {4, 196}

\bibitem[\protect\citeauthoryear{{Freedman} et~al.,}{{Freedman}
  et~al.}{2020}]{TRGB}
{Freedman} W.~L.,  et~al., 2020, \mn@doi [\apj] {10.3847/1538-4357/ab7339},
  \href {https://ui.adsabs.harvard.edu/abs/2020ApJ...891...57F} {891, 57}

\bibitem[\protect\citeauthoryear{{Heymans} et~al.,}{{Heymans}
  et~al.}{2020}]{KIDS1000}
{Heymans} C.,  et~al., 2020, arXiv e-prints, \href
  {https://ui.adsabs.harvard.edu/abs/2020arXiv200715632H} {p. arXiv:2007.15632}

\bibitem[\protect\citeauthoryear{{Hildebrandt} et~al.,}{{Hildebrandt}
  et~al.}{2017}]{KIDS450}
{Hildebrandt} H.,  et~al., 2017, \mn@doi [\mnras] {10.1093/mnras/stw2805},
  \href {https://ui.adsabs.harvard.edu/abs/2017MNRAS.465.1454H} {465, 1454}

\bibitem[\protect\citeauthoryear{{Hill}, {McDonough}, {Toomey}  \&
  {Alexander}}{{Hill} et~al.}{2020}]{EDE2}
{Hill} J.~C.,  {McDonough} E.,  {Toomey} M.~W.,   {Alexander} S.,  2020, arXiv
  e-prints, \href {https://ui.adsabs.harvard.edu/abs/2020arXiv200307355H} {p.
  arXiv:2003.07355}

\bibitem[\protect\citeauthoryear{{Howlett}, {Ross}, {Samushia}, {Percival}  \&
  {Manera}}{{Howlett} et~al.}{2015}]{HowlettBAOMGS}
{Howlett} C.,  {Ross} A.~J.,  {Samushia} L.,  {Percival} W.~J.,   {Manera} M.,
  2015, \mn@doi [\mnras] {10.1093/mnras/stu2693}, \href
  {https://ui.adsabs.harvard.edu/abs/2015MNRAS.449..848H} {449, 848}

\bibitem[\protect\citeauthoryear{Joudaki, Kaplinghat, Keeley  \&
  Kirkby}{Joudaki et~al.}{2018}]{KeeleyGP1}
Joudaki S.,  Kaplinghat M.,  Keeley R.~E.,   Kirkby D.,  2018, \mn@doi [Phys.
  Rev. D] {10.1103/PhysRevD.97.123501}, 97, 123501

\bibitem[\protect\citeauthoryear{{Keeley}, {Joudaki}, {Kaplinghat}  \&
  {Kirkby}}{{Keeley} et~al.}{2019}]{Keeley19GP2}
{Keeley} R.~E.,  {Joudaki} S.,  {Kaplinghat} M.,   {Kirkby} D.,  2019, \mn@doi
  [\jcap] {10.1088/1475-7516/2019/12/035}, \href
  {https://ui.adsabs.harvard.edu/abs/2019JCAP...12..035K} {2019, 035}

\bibitem[\protect\citeauthoryear{{Keeley}, {Shafieloo}, {L'Huillier}  \&
  {Linder}}{{Keeley} et~al.}{2020}]{KeeleyGP3}
{Keeley} R.~E.,  {Shafieloo} A.,  {L'Huillier} B.,   {Linder} E.~V.,  2020,
  \mn@doi [\mnras] {10.1093/mnras/stz3304}, \href
  {https://ui.adsabs.harvard.edu/abs/2020MNRAS.491.3983K} {491, 3983}

\bibitem[\protect\citeauthoryear{{Kim}, {Kang}  \& {Lee}}{{Kim}
  et~al.}{2019}]{kim-etal19}
{Kim} Y.-L.,  {Kang} Y.,   {Lee} Y.-W.,  2019, \mn@doi [Journal of Korean
  Astronomical Society] {10.5303/JKAS.2019.52.5.181}, \href
  {https://ui.adsabs.harvard.edu/abs/2019JKAS...52..181K} {52, 181}

\bibitem[\protect\citeauthoryear{Kirkby \& Keeley}{Kirkby \&
  Keeley}{2017}]{gphistdoi}
Kirkby D.,  Keeley R.~E.,  2017, Cosmological expansion history inference using
  Gaussian processes, \mn@doi{10.5281/zenodo.999564}

\bibitem[\protect\citeauthoryear{{Knox} \& {Millea}}{{Knox} \&
  {Millea}}{2020}]{HubbleHunter}
{Knox} L.,  {Millea} M.,  2020, \mn@doi [\prd] {10.1103/PhysRevD.101.043533},
  \href {https://ui.adsabs.harvard.edu/abs/2020PhRvD.101d3533K} {101, 043533}

\bibitem[\protect\citeauthoryear{{L'Huillier}, {Shafieloo}, {Linder}  \&
  {Kim}}{{L'Huillier} et~al.}{2019}]{BenShafKimLind2019}
{L'Huillier} B.,  {Shafieloo} A.,  {Linder} E.~V.,   {Kim} A.~G.,  2019,
  \mn@doi [\mnras] {10.1093/mnras/stz589}, \href
  {https://ui.adsabs.harvard.edu/abs/2019MNRAS.485.2783L} {485, 2783}

\bibitem[\protect\citeauthoryear{{Li} \& {Shafieloo}}{{Li} \&
  {Shafieloo}}{2019}]{PEDE}
{Li} X.,  {Shafieloo} A.,  2019, \mn@doi [\apjl] {10.3847/2041-8213/ab3e09},
  \href {https://ui.adsabs.harvard.edu/abs/2019ApJ...883L...3L} {883, L3}

\bibitem[\protect\citeauthoryear{{Li} \& {Shafieloo}}{{Li} \&
  {Shafieloo}}{2020}]{GEDE}
{Li} X.,  {Shafieloo} A.,  2020, arXiv e-prints, \href
  {https://ui.adsabs.harvard.edu/abs/2020arXiv200105103L} {p. arXiv:2001.05103}

\bibitem[\protect\citeauthoryear{{Liao}, {Shafieloo}, {Keeley}  \&
  {Linder}}{{Liao} et~al.}{2020}]{keeleykai}
{Liao} K.,  {Shafieloo} A.,  {Keeley} R.~E.,   {Linder} E.~V.,  2020, \mn@doi
  [\apjl] {10.3847/2041-8213/ab8dbb}, \href
  {https://ui.adsabs.harvard.edu/abs/2020ApJ...895L..29L} {895, L29}

\bibitem[\protect\citeauthoryear{{Planck Collaboration} et~al.,}{{Planck
  Collaboration} et~al.}{2018}]{Planck18Cosmo}
{Planck Collaboration} et~al., 2018, arXiv e-prints, \href
  {https://ui.adsabs.harvard.edu/abs/2018arXiv180706209P} {p. arXiv:1807.06209}

\bibitem[\protect\citeauthoryear{{Poulin}, {Smith}, {Karwal}  \&
  {Kamionkowski}}{{Poulin} et~al.}{2019}]{EDE}
{Poulin} V.,  {Smith} T.~L.,  {Karwal} T.,   {Kamionkowski} M.,  2019, \mn@doi
  [\prl] {10.1103/PhysRevLett.122.221301}, \href
  {https://ui.adsabs.harvard.edu/abs/2019PhRvL.122v1301P} {122, 221301}

\bibitem[\protect\citeauthoryear{Rasmussen \& Williams}{Rasmussen \&
  Williams}{2006}]{Rasmussen}
Rasmussen C.~E.,  Williams C. K.~I.,  2006, Gaussian Processes for Machine
  Learning.
the MIT Press

\bibitem[\protect\citeauthoryear{{Riess}, {Casertano}, {Yuan}, {Macri}  \&
  {Scolnic}}{{Riess} et~al.}{2019}]{Riess2019}
{Riess} A.~G.,  {Casertano} S.,  {Yuan} W.,  {Macri} L.~M.,   {Scolnic} D.,
  2019, \mn@doi [\apj] {10.3847/1538-4357/ab1422}, \href
  {https://ui.adsabs.harvard.edu/abs/2019ApJ...876...85R} {876, 85}

\bibitem[\protect\citeauthoryear{{Ross}, {Samushia}, {Howlett}, {Percival},
  {Burden}  \& {Manera}}{{Ross} et~al.}{2015}]{RossBAOMGS}
{Ross} A.~J.,  {Samushia} L.,  {Howlett} C.,  {Percival} W.~J.,  {Burden} A.,
  {Manera} M.,  2015, \mn@doi [\mnras] {10.1093/mnras/stv154}, \href
  {https://ui.adsabs.harvard.edu/abs/2015MNRAS.449..835R} {449, 835}

\bibitem[\protect\citeauthoryear{{Scolnic} et~al.,}{{Scolnic}
  et~al.}{2018}]{Pantheon}
{Scolnic} D.~M.,  et~al., 2018, \mn@doi [\apj] {10.3847/1538-4357/aab9bb},
  \href {https://ui.adsabs.harvard.edu/\#abs/2018ApJ...859..101S} {859, 101}

\bibitem[\protect\citeauthoryear{{Shafieloo}, {Kim}  \& {Linder}}{{Shafieloo}
  et~al.}{2012}]{ShafKimLind2012}
{Shafieloo} A.,  {Kim} A.~G.,   {Linder} E.~V.,  2012, \mn@doi [\prd]
  {10.1103/PhysRevD.85.123530}, \href
  {https://ui.adsabs.harvard.edu/abs/2012PhRvD..85l3530S} {85, 123530}

\bibitem[\protect\citeauthoryear{Shafieloo, Kim  \& Linder}{Shafieloo
  et~al.}{2013}]{ShafKimLind2013}
Shafieloo A.,  Kim A.~G.,   Linder E.~V.,  2013, \mn@doi [Phys. Rev. D]
  {10.1103/PhysRevD.87.023520}, 87, 023520

\bibitem[\protect\citeauthoryear{{Wong} et~al.,}{{Wong} et~al.}{2020}]{H0LiCOW}
{Wong} K.~C.,  et~al., 2020, \mn@doi [\mnras] {10.1093/mnras/stz3094}, \href
  {https://ui.adsabs.harvard.edu/abs/2020MNRAS.tmp.1661W} {}

\bibitem[\protect\citeauthoryear{{du Mas des Bourboux} et~al.,}{{du Mas des
  Bourboux} et~al.}{2017}]{SDSSLya}
{du Mas des Bourboux} H.,  et~al., 2017, \mn@doi [\aap]
  {10.1051/0004-6361/201731731}, \href
  {https://ui.adsabs.harvard.edu/abs/2017A&A...608A.130D} {608, A130}

\bibitem[\protect\citeauthoryear{{du Mas des Bourboux} et~al.,}{{du Mas des
  Bourboux} et~al.}{2020}]{eBOSSLya}
{du Mas des Bourboux} H.,  et~al., 2020, arXiv e-prints, \href
  {https://ui.adsabs.harvard.edu/abs/2020arXiv200708995D} {p. arXiv:2007.08995}

\bibitem[\protect\citeauthoryear{{eBOSS Collaboration} et~al.,}{{eBOSS
  Collaboration} et~al.}{2020}]{eBOSSCosmo}
{eBOSS Collaboration} et~al., 2020, arXiv e-prints, \href
  {https://ui.adsabs.harvard.edu/abs/2020arXiv200708991E} {p. arXiv:2007.08991}

\makeatother
\end{thebibliography}




\bsp	
\label{lastpage}
\end{document}